# Teleology-Driven Affective Computing:
# A Causal Framework for Sustained Well-Being

Bin Yin#, Chong-Yi Liu#, Liya Fu and Jinkun Zhang

*Abstract —* Affective computing has advanced significantly over the past two decades, achieving remarkable progress in emotion recognition and generation. However, current approaches remain largely focused on short-term pattern recognition, lacking a comprehensive theoretical framework to guide affective agents in aligning with long-term human well-being. To address this gap, we propose a teleology-driven affective computing framework, which unifies major emotion theories — basic emotion, appraisal, and constructivist approaches — under the premise that affect is an adaptive, goal-directed process that facilitates survival and development. Building on this foundation, our framework emphasizes the alignment of agent responses with personal/individual and group/collective well-being over extended timescales. We advocate for developing a large-scale "dataverse" of personal affective events, capturing the longitudinal interplay between beliefs, goals, actions, and outcomes using real-world experience sampling and immersive virtual reality. By leveraging causal modeling, this "dataverse" enables AI systems to infer individuals' unique affective concerns and provide tailored interventions that support sustained well-being. Furthermore, we introduce a meta-reinforcement learning paradigm to train affective agents in simulated environments. This allows them to dynamically adapt to evolving affective concerns and balance hierarchical goals—ranging from immediate affective concerns to long-term self-actualization. We call for a shift from traditional modeling based on statistical correlations to causal reasoning, enhancing agents' ability to predict and proactively respond to emotional challenges. Ultimately, this framework offers a conceptual foundation for developing adaptive, personalized, and ethically aligned affective computing systems that promote meaningful human-AI interactions and long-term societal well-being.

## 1. INTRODUCTION

THE scientific community, long emphasizing rational thought, logical reasoning, operationalization, and reproducibility, has often regarded emotions with a certain detachment. Consequently, when Picard [1] first proposed the concept of affective computing, it garnered limited attention. However, addressing certain engineering challenges inevitably involves engaging with emotions, often considered "irrational." For instance, early computer vision systems were largely modeled after the structure and function of the human visual cortex [2]. Yet, creating such systems was not merely about detecting high-contrast lines or distinguishing apples from pears; it required identifying elements of interest and adapting to shifting priorities automatically — tasks that no purely engineering-based approach could achieve [3]. Marvin Minsky emphasized that emotions are not the antithesis of rationality but are critical mechanisms enabling humans to make efficient decisions in complex environments [4]. He viewed emotions as a strategic selection system governing thoughts and actions, determining when to persist, abandon, or switch goals. This perspective aligns closely with the core objectives of affective computing, which aims to endow computational systems with similar strategic adaptability, a foundational step toward intelligent human-computer interaction.

The value of affective computing has been validated in various artificial intelligence (AI) applications, such as intelligent assistants, social robots, virtual reality, and augmented reality systems. The capacity for emotional understanding and responsiveness significantly enhances user experiences. Beyond enabling more natural and human-like interactions, affective computing also plays vital roles in healthcare, education, and psychotherapy [5]. Driven by this expanded understanding of emotional functionality, an increasing number of engineers and computer scientists have been drawn to this previously "niche" field. They provide a variety of information technology tools and engineering capabilities to create a computing system that can accurately perceive, recognize, understand and respond to human emotions based on behavioral cues and contextual situations [5]. As a result, affective computing has evolved into an interdisciplinary domain, integrating insights from psychology, computer science, information technology, mechanical engineering, and bioengineering [5], [6]. Researchers aim to create intelligent agents that better understand human emotional needs [1], fostering harmonious human-computer interactions and advancing user-centered AI [7], [8], [9].

Current research in affective computing primarily focuses on two areas: emotional analysis [10], [11] and the generation and expression of emotions [12], [13]. Emotional analysis involves modeling and recognizing multidimensional emotional signals

---
# These authors contributed equally to this work and are co-first authors.
All authors are affiliated with the School of Psychology, Fujian Normal University, Fuzhou, Fujian 350117, China.
Corresponding author: Bin Yin, Ph.D. (Email: byin@fjnu.edu.cn; luckbin@163.com; ORCID: 0000-0003-1105-8905).

collected from text, speech, and visual inputs [14], while emotional generation leverages generative models to simulate human-like emotional expressions [15]. Recent advancements in deep learning [11] and the creation of large datasets [16] have significantly enhanced emotional analysis. For example, in text-based emotional analysis, a critical research challenge is classifying sentiments (e.g., positive or negative) from textual data, such as product reviews. Domain shift problems often arise when training and testing data originate from different distributions, leading to performance degradation. To address this, Ganin et al., [17] introduced Domain Adversarial Neural Networks (DANN) to bridge distributional gaps between source and target domains.

Despite achieving high accuracy in emotion recognition tasks through machine learning models trained on large datasets, these models still lack genuine emotional understanding. They often fail to infer the causes of emotions, predict behavioral outcomes, or hypothesize how interventions might alter emotional trajectories. In the realm of emotional generation, large language models (LLMs) like ChatGPT can produce word-for-word responses resembling human emotional expressions through specific training methods. However, true empathy and emotional concern remain beyond the capabilities of LLMs [18]. While these models can simulate forms of empathetic and affectionate expressions, they do not replicate the essence or underlying generative processes of emotional experiences. For humans, identifying others' emotional states and expressing one's own emotions is a continuous and integrated process. However, whether built on discrete [19], [20] or dimensional models of emotion [21], [22], [23], existing algorithms for emotion recognition often fail to guide affective AI systems on actionable responses. This highlights the absence of a comprehensive framework in affective computing.

The complexity and diversity of affective phenomena pose challenges for traditional reductionist methods to fully elucidate their underlying mechanisms. Fundamental questions in affective computing include: What principles govern emotional interactions? How are emotional states formed and developed through individual-environment interactions? How to effectively distinguish the different triggers of affective states, such as grief caused by the loss of a loved one and sadness triggered by a decline in academic performance? After recognizing emotions, how should AI systems respond appropriately to support effective emotional interaction? Addressing these questions requires a clear theoretical framework and robust technical solutions to bridge the gap between emotion recognition and decision-making. Teleology offers a systematic perspective for understanding the functional roles of emotions in adapting to environments and achieving goals [24]. By incorporating teleological principles, we aim to redefine the design of affective agents. This paper introduces a teleology-based research framework for affective computing, encompassing theoretical foundations, interaction principles, dataset development, and algorithmic modeling. By conceptualizing affect as adaptive and goal-oriented products, our objective is to surpass traditional pattern recognition methods and provide a systematic perspective for advancing the field.

The remainder of this paper is organized as follows: Section 2 examines major contemporary emotion theories from a teleological perspective. Section 3 reviews the state of research in affective computing, focusing on mainstream computational models and their theoretical assumptions. Section 4 proposes interaction principles for affective agents based on teleology and explores the construction of corresponding datasets and algorithmic models, culminating in a comprehensive teleology-based affective computing framework. Section 5 summarizes and provides conclusions.

## 2. A Teleological Perspective on Major Emotion Theories

Contemporary emotion theories have yet to reach a consensus on fundamental assumptions, with even the basic definition of "emotion" remaining a contentious issue [25], [26], [27], [28], [29], [30]. These theoretical paradigms propose various perspectives, including basic emotion theories, appraisal theories, and constructivist approaches (both psychological and social) [31]. While these theories offer unique perspectives, long-standing criticisms and misunderstandings have hindered the full development and application of their core ideas. This has left the field of affective computing without a unified theoretical framework. Just as the Industrial Revolution relied on the foundational principles of Newtonian physics to drive technological progress, achieving the goal of creating genuine affective agents requires addressing the foundational gaps in psychological theories of emotion and integrating these insights into computational systems. Teleology provides a powerful integrative framework by emphasizing affects as functional tools that organisms use to adapt to their environments and achieve goals. This perspective bridges the divides between basic emotion theories, appraisal theories, and constructivist approaches, underscoring the core role of affects in helping individuals adapt and achieve their objectives.

When applied to psychology, teleology posits that mental constructs exist to serve adaptive or functional purposes [32]. In studying affective phenomena, this perspective prompts us to consider why humans evolved affects. To serve evolutionary purposes, affects must be linked to functional components in the physical world [33]. Specifically, affects can be defined as psychological mechanisms supporting life [25], [29] and naturally extend the observation of affective phenomena from humans to other animals [34], [35], [36], [37], [38], [39], [40], [41], [42]. From this perspective, affect serve as internal states that mediate between causes and outcomes of actions. Their functional significance can be demonstrated through operant and reinforcement processes, revealing remarkable conservation across species. Despite criticism tying consciousness tightly to affects and questioning the study of affects in animals unable to self-report, neuroscience research increasingly reveals convergent neural mechanisms underlying affective responses across species [37], [43], [44], [45]. Panksepp [44] introduced the concept of "affective neuroscience," integrating the shared functions of affective systems across species and highlighting affects as foundational to the development of complex cognitive abilities.

The teleological approach provides a psychological rationale for explaining behavior, focusing on biological needs or goals [32]. For instance, fear (an emotional response) triggers behavioral actions (such as freezing or fleeing) aimed at protecting the organism from potential harm. Each action is goal-driven during interactions with the external environment. In contrast, mechanisms based purely on mathematics, probability, geometry, or logic, which disregard physical implementation factors, do not align with the goal-oriented paradigm that underlies biological intelligence [46]. At a broader biological level, maintaining self-continuity and coherence serves as the overarching principle guiding all actions [24]. This principle ensures that biological systems maintain homeostasis within specific ranges [47]. The fundamental biological goals of securing comfort and propagating genes underpin a wide range of affective phenomena, from simple interests and moods to more complex desires, ethics, and aesthetics [44]. Cross-species research provides evidence for the evolutionary continuity of emotional systems, which act as a universal currency in decision-making: pleasure motivates behavior persistence, while discomfort drives avoidance [48].

Modern theories of emotion emphasize their adaptive functions. Basic emotion theories agree that distinct functional emotions evolved to address specific adaptive challenges — recurring threats or opportunities related to survival and health [44], [47], [49]. For example, Ekman [50] argued that emotional processes coordinate physiological, non-verbal, attentional, cognitive, motivational, sensory, and behavioral responses when triggered by specific contexts. Appraisal theorists similarly recognize the adaptive nature of emotions, defining them as mechanisms that drive organisms to respond appropriately to environmental changes. Arnold [51] described emotions as approach responses to beneficial stimuli or avoidance responses to harmful ones. This evaluation process continuously monitors the relationship between an organism and its "comfort zone", integrating multiple dimensions such as needs, goals, and values [52]. Thus, even in identical situations, individuals may generate different emotional responses based on their unique appraisals. Constructivist approaches view emotions as brain-generated constructs shaped by combinations of basic sensory inputs, concepts, and categories, influenced by both cognitive processes and biological factors. Emotional expressions and physiological features are closely tied to past experiences, exhibiting flexibility and context dependence. For instance, a smile's muscle movements might convey anger in one context but friendliness or joy in another.

The classification principles of emotions reflect differing extensions of teleological assumptions in emotional theories. Scientists debate whether emotions should be described along dimensions like valence and arousal or as discrete entities [53]. Basic emotion theories typically adopt a discrete view, dividing emotions into categories defined by specific emotional programs. Each category represents an organism's relationship with its environment, such as fear protecting against harm or disgust rejecting harmful substances [50]. In contrast, appraisal theories argue for dimensional constructs, suggesting that emotions arise from evaluative patterns along multiple dimensions. These dimensions combine in diverse ways to produce a broad spectrum of emotional states. For appraisal theorists, emotions are valenced according to their alignment with goals or goal states. Positive emotions arise when goals are achieved, while negative emotions reflect failures. An organism's primary task is not to process information but to fulfill its goals, necessitating continuous cycles of actions that serve these goals [54], [55]. Appraisal theory conceptualizes goals as the individual's well-being, with the appraisal process serving to detect and evaluate the importance of the environment's impact on well-being [56]. Within this framework, the appraisal process assesses the influence of the environment on an individual's well-being, incorporating multidimensional factors such as needs, beliefs, and attachments [52], [56], [57], [58]. It is crucial to recognize that these concerns precede specific emotions and exist independently of them, functioning as an emotional state akin to desire [59].

While constructivism recognizes the utility of emotion categories, it emphasizes that valence and arousal are central features that vary depending on the context and the individual's experiences. Emotional instances emerge as events occurring in specific contexts. When the brain reconstructs past events similar to the present, it generates categories containing potential future emotional instances [27]. These categories encapsulate abstract psychological features such as goals, values, and threats, emphasizing the importance of concerns in the emotional generation process.

In summary, although differing metaphysical and mechanistic assumptions underpin these theories, teleological principles — focusing on goal-directed behaviors and adaptive functions — offer a way to unify them into a cohesive framework. Schiller et al. [24] proposed the concept of the "Human Affectome," a framework integrating diverse assumptions about emotions into a unified model. This model views affects as dynamic phenomena emerging from the interactions between organisms and their environments, encompassing both concerns (factors influencing affective experiences) and features (adaptive changes such as valence and arousal). This framework provides a novel perspective and integrative strategies for advancing research in affective computing.

## 3. Current Research in Affective Computing

Current research in affective computing can be broadly divided into two main domains: emotional analysis and emotional generation/expression. The former focuses on modeling and recognizing patterns in external emotional expressions (such as facial expressions, vocal tones, and movements) and physiological signals (e.g., heart rate, blood pressure, pulse, and body temperature). The latter aims to develop affective agents capable of generating emotions and assessing possible user emotional responses by calculating events and evaluating outcomes [61]. This section will analyze mainstream computational models and research achievements in these two domains, as well as evaluate the theoretical assumptions underpinning them.

*3.1 Emotional Analysis － Modeling and Recognizing Emotional Representations*

Living organisms consistently use emotional expressions to communicate with others, even across species. For humans, such expressions aid in coordinating personal, interpersonal, and social behaviors. One of the key goals of affective computing is to "understand" the underlying emotional states of interaction partners. Humans typically express emotions through a combination of verbal communication, vocal intonation, facial expressions, and body language. Researchers have attempted to infer users' emotional states using single-modal physical data, with studies focusing on text sentiment analysis, speech emotion recognition, and visual emotion recognition (for detailed reviews of these three areas, see [9]). Traditional single-modal emotion recognition typically relies on a process called "feature engineering" [62], where relevant emotion-related features are extracted from raw data and input into predictive models that output corresponding emotional states. Deep neural networks, by aggregating activations from interconnected layers of processing units (neurons), can automatically extract useful features from raw data [63], enabling end-to-end emotional analysis. However, pattern recognition based on external emotional expressions is not always reliable. Social norms or individual concerns about judgment can lead people to unconsciously or deliberately mask their true emotional states, reducing the accuracy of single-modal emotion recognition methods.

In uncovering hidden emotions, micro-expression recognition (MER) in visual emotion analysis often performs better [64], [65], [66]. Ekman's research showed that micro-expressions, as a specific type of facial expression, are typically involuntarily displayed when individuals attempt to hide their true feelings, reflecting spontaneous facial movements in response to emotional stimuli [67], [68], [69]. Thus, micro-expressions provide a valuable source of information for decoding authentic emotional states. Like general facial expression recognition, MER involves categorizing facial images or sequences into corresponding emotion classes. However, detecting these fleeting, subtle facial movements is often more challenging [64]. Early MER approaches also relied on manually crafted facial features. Tools such as the Facial Action Coding System (FACS) developed by Ekman [70] and the Micro-Expression Training Tool (METT) provided researchers with essential prior knowledge for encoding features (for comprehensive reviews of MER's historical evolution and technological advancements, see [71], [72], [73], [74]). In recent years, more advanced models, such as 3D Convolutional Neural Networks (3D CNNs) and Recurrent Neural Networks (RNNs), have been used to capture the spatiotemporal characteristics of micro-expressions (e.g., [75], [76], [77], [78], [79], [80], [81], [82]). Despite significant progress achieved using deep learning-based MER, challenges remain, particularly concerning datasets. Guoying Zhao and collaborators [64], [83] found that although numerous micro-expression datasets have been developed, most remain limited in size, containing only a few hundred samples. Additionally, these datasets often suffer from imbalanced emotion categories and are predominantly collected in laboratory settings, which limits the generalizability of MER models to real-world environments.

Physiological signals (e.g., EEG, skin conductance, and ECG) also represent emotion-related indicators that are difficult to consciously control, as they reflect central and autonomic nervous system activities associated with emotional states [84], [85]. Picard and her team utilized multi-channel signals from the autonomic nervous system to identify human emotional states. They developed various wearable devices [86], [87] and non-contact devices [88], [89] capable of recording physiological signals such as skin conductance, heart rate, and physical activity in real-life scenarios. These physiological measurements provide objective data to identify risk factors for high stress and poor mental health [90], [91], [92], [93]. Moreover, electroencephalography (EEG) has become an essential signal for recording central nervous system activity due to its non-invasive, fast, portable, and cost-effective nature. EEG signals are widely used in fields such as entertainment, virtual reality, and e-healthcare for identifying emotional states [94], [95]. While EEG signals directly reflect the brain's electrophysiological activity, linking them to the central nervous mechanisms of emotional states [96], practical challenges remain in model transferability and generalization. In EEG emotion recognition research, researchers often address individual differences within the same dataset to construct models generalizable to all individuals [97], [98], [99], [100]). However, variability in devices, experimental designs, and stimuli across datasets necessitates further exploration of how knowledge trained on one dataset can be transferred to others [96].

When external environments become overly complex, single-modal signals are often affected by noise, resulting in reduced emotion recognition performance. Considering that humans express emotions through multiple modalities rather than relying on a single channel, research has increasingly focused on multimodal emotion recognition (e.g., [14], [101], [102], [103], [104]). Similar to single-modal approaches, multimodal emotion recognition involves extracting features from raw data across different modalities and training classifiers to predict emotional states. However, inherent issues arise from asynchronous sampling rates across modalities, leading to misaligned data during fusion [103], [105]. To address this, Tsai et al., [103] developed a model based on the Transformer framework that facilitates cross-modal interactions. The multi-head attention module in this framework allows for directional pairwise cross-modal attention and further focuses on interactions between multimodal sequences across varying time steps, achieving alignment and fusion of information. Additionally, recent studies have focused on addressing feature space distribution gaps caused by multimodal signal heterogeneity (e.g., [101], [102], [104]). In general, multimodal emotion analysis requires selecting appropriate single-modal emotion data and determining a fusion strategy, which may occur at the feature level, decision level, model level, or through hybrid approaches [9], [106]. The success of multimodal emotion analysis systems lies in the careful choice of data and fusion strategy [73], which often outperform single-modal recognition systems [107].

*3.2 Modeling Emotional Generation Processes*

In the domain of modeling emotional generation, appraisal theories have become dominant. Many computational models based on appraisal theories have been widely applied in human-computer interaction systems, particularly for generating emotional expressions in real-time interactive characters (e.g., [108], [109]). The central assumption of appraisal theories is that individuals continuously monitor and evaluate changes in the environment that are relevant to their concerns, with emotions arising from these evaluations [110], [111]. Lazarus [52] introduced the concept of the "person-environment relationship" to describe the adaptive processes between agents and their environments. This relationship allows agents to derive the significance of external events relative to their beliefs, desires, and intentions, thereby generating or inferring emotional responses. Some computational models do not explicitly represent these relationships but still support the derivation of event meanings to produce emotional reactions. For instance, the EMA model (named in honor of Richard Lazarus's book *Emotion and Adaptation*) [112] constructs causal explanations of the current situation, incorporating beliefs, desires, and intentions. These causal explanations generate appraisal frames for significant events, which in turn yield emotional responses based on the outcomes of these appraisals. Other approaches have employed Bayesian inference [113], [114], [115] or Partially Observable Markov Decision Processes (POMDP) [116] to model this derivation process. More recently, the Belief-Desire-Intention (BDI) software architecture, which simulates human cognitive processes [117], has been increasingly used to design affective agents (e.g., [118], [119], [120], [121]). These BDI-based agents can not only model the process of emotion elicitation (i.e., how emotions are triggered based on the agent's beliefs, desires, and intentions) but also simulate the impact of emotions on cognitive processes and action choices [122].

The cognitive appraisal theory proposed by Ortony, Clore, and Collins (OCC), [123] further refines the triggers for emotional generation by distinguishing between events, actions by others, and objects as different types of stimuli. Based on these triggers, the theory provides a new classification scheme for emotions and establishes structural relationships between specific appraisal variables and emotional labels. For example, it specifies how combinations of appraisal variables can elicit emotions such as shame [124]. Due to its structured nature, the OCC model is considered a "computationally friendly" framework for modeling emotion generation [61]. Dias et al., [125] built the Fearnot AffecTIve Mind Architecture (FAtiMA) based on the OCC model, which integrates planning capabilities into affective agents. A key component of this architecture, OCCAffectDerivation, generates emotions based on appraisal variables derived from the OCC model, such as desirability, desirability-for-others, and praiseworthiness. For example, if an event has a positive desirability value exceeding the pre-defined threshold for joy, the model generates the emotion of joy. Conversely, events with negative desirability may result in distress. Other computational models also employ the appraisal variables proposed in the OCC theory, including AR [126], EM [108], FLAME [127], ALMA [128] and GAMYGDALA [109].

However, as Hudlicka [61] noted, the highly structured nature of the OCC model is a double-edged sword. On the one hand, its detailed specifications provide a clear and systematic guideline for computational representation of emotions. On the other hand, the fine-grained descriptions of specific structures may not always be necessary, as certain models might not need to differentiate between events, actions by others, or objects to derive emotions.

In addition to the OCC theory, other cognitive appraisal theorists have proposed their own sets of appraisal variables that intelligent agents can use to generate distinct emotional responses, e.g., Frijda [129], Roseman [130], Scherer [131], C. A. Smith and Ellsworth [132]. These variables include novelty, intrinsic valence, certainty, goal conduciveness, agency, control, and compatibility with personal or social standards [133]. While these variables are less structurally detailed compared to the OCC model, they provide a unified framework for emotion computation, where specific emotions can be represented as vectors of appraisal variables. Scherer's proposed variables [58], [110], [131] have been adopted in computational models such as WASABI [134] and PEACTIDM [135]. These theories clarify the mapping between appraisal variables and emotional states, modeling the appraisal process as the cause of emotion generation. Once the pattern of appraisal variables is determined, corresponding emotional states can be derived through simple if-then rules, often represented by discrete emotion labels [136], [137].

Early computational appraisal models often employed symbolic rule-based systems to map appraisal variables to specific emotional states (e.g., [108], [127], [128], [134], [135]). However, with the rapid development of deep learning and neural network technologies [63], [138], connectionist approaches have gained momentum. Researchers now use neural networks composed of numerous simple processing units to simulate the appraisal process [139], [140], [141]. Deep learning-based affective computing has become a dominant trend, as the strong fitting capabilities of deep neural networks allow them to model complex relationships between data without relying on explicitly programmed rules. Ong et al., [140] applied the paradigm of deep probabilistic programming to affective computing, combining data-driven (deep learning) and theory-driven (probabilistic programming) approaches to model uncertainties in the emotional generation process. Additionally, LLMs have introduced new methods for simulating the appraisal process, as they can generate semantically meaningful outputs based on extensive text datasets that may implicitly represent certain human psychological processes [142], [143]. In practice, Wei et al., [144] demonstrated that the performance of LLMs depends not only on the model itself but also on the quality of prompts. They proposed the "Chain-of-Thought" prompting concept, which uses step-by-step reasoning examples to guide LLMs, significantly enhancing their reasoning capabilities. Croissant et al. [145] applied this paradigm to simulate the emotional appraisal process. By integrating contextual information and role-playing elements into a memory system, their approach allowed LLMs to evaluate situations and generate or infer emotions based on appraisal prompts. However, while LLMs trained with deep learning can provide contextually appropriate responses based on semantic understanding [142], [146], [147], they do not simulate the genuine emotional generation process. Instead, they produce statistically plausible responses based on input data. Consequently, the performance of these deep learning-based appraisal models remains limited by the quality of training data and the inherent statistical properties of the models [148].

*3.3 Limitations of Current Affective Computing Models and Future Directions*

Both single-modal and multimodal emotion recognition approaches share a common foundational assumption: internal emotional states manifest in observable external and physiological signals. These measurable signals are then mapped back to corresponding emotional states using discrete or dimensional models. Discrete models classify emotions into a finite set of categories. Among them, Ekman's basic emotion theory and its variants have been widely applied in emotion recognition research. Ekman proposed that certain emotions (e.g., anger, disgust, fear, happiness, sadness, and surprise) are universally present across cultures and assumed that each basic emotion corresponds to distinct response components (e.g., facial expressions and physiological reactions) that tend to co-occur during emotional episodes [50].

However, empirical studies have shown that even when researchers use the same emotional labels, the described emotional states often differ [149], [150]. Additionally, current neuroimaging techniques have not provided evidence supporting the existence of distinct "neural fingerprints" for each basic emotion [151]. While Ekman acknowledged the existence of variants within each basic emotion family and the blending of emotions at specific moments, he also maintained that the boundaries between basic emotion categories should remain distinct. Neuroscientific research increasingly highlights that complex psychological phenomena, including emotions, are mediated by distributed neural activation networks [152]. This suggests that, from a neurophysiological perspective, clear boundaries between different categories of basic emotions may not exist [60]. In fact, contemporary psychology research places greater emphasis on the functional labels of basic emotions. Many studies suggest functionally distinguishing basic emotions; for instance, "disgust" can be further categorized into emotional responses to contamination versus destruction, reflecting different adaptive problems [149], [153], In animal studies, researchers have linked emotional elicitation contexts, neural circuits, and observable behaviors, documenting responses resembling basic emotions. However, researchers often avoid colloquial emotional labels such as "joy" or "fear" and prefer motivational behavioral labels like "wanting" or "defense" to describe these reactions [44], [154], [155].

The computational models in emotion recognition have yet to incorporate the functional characteristics of emotions fully. Even models based on dimensional theories have similar limitations (e.g., [156], [157]). While these techniques are valuable for applications such as public opinion monitoring, sentiment analysis, and health management, they often fall short of achieving the ultimate goal of creating affective agents capable of truly understanding user emotions and providing appropriate responses. Key questions remain unanswered: **What constitutes a "correct response" in affective interaction systems? What defines a "natural and smooth" interaction?** Answering these questions requires a teleological perspective to understand the entire process of emotion generation and its functional significance. Currently, emotional response systems typically follow a default approach: reading human emotional cues, categorizing emotions, and responding based on predefined patterns. In these designs, the complexity and diversity of emotions are often reduced to fixed categories and preset reaction patterns. While convenient to implement, these designs overlook the deeper purposiveness and dynamism of emotions, making them less adaptable and realistic in long-term interactions. For instance, Terzis et al. [158] designed an emotion feedback system employing parallel and reactive empathy to manage learners' six emotional states (happiness, anger, sadness, surprise, fear, and disgust). When a learner displayed fear, the system first mirrored the emotion through parallel empathy, using fearful facial expressions and vocal tones, and then shifted to joyful expressions and tones (reactive empathy) to alleviate the learner's fear and encourage continued efforts. While practical, this approach has inherent flaws. Firstly, the psychological validity of such categorical classifications remains debated. Secondly, the lack of understanding of the teleological principles behind emotions limits the system's ability to influence and alter users' emotional states effectively. In psychology, empathy involves perceiving the inner subjective world of another person, understanding their sources of distress or sorrow, and conveying meaningful responses. Emotional generation is thus a constructive process, deeply individualistic [27]. Predefined uniform responses cannot adequately simulate genuine emotional responses.

Primarily focuses on individual-level dynamics, appraisal theory enables agents to generate or understand emotions based on their evaluations of psychological events concerning their own or others' concerns. It addresses key questions: How are specific emotions mapped from stimulus events? Are these mappings direct, or do they involve intermediate appraisal processes (e.g., novelty, desirability, or agency)? How do internal stimuli (e.g., memories or anticipated events) interact with external stimuli (reflecting contextual features) to jointly influence emotion generation? Answering these questions allows continuous comparison between environmental changes and individual affective concerns, providing opportunities to operationalize teleological principles in emotional machines. Accordingly, some computational appraisal models enable developers to set specific affective concerns for affective agents [109]. These concerns allow agents to perform personalized actions based on their internal goals. However, such hand-designed affective agents can enhance user experiences only in narrowly defined scenarios, such as gaming, and often fail to respond to the affective concerns of individuals during complex interactions.

Moreover, models trained on large datasets and deep learning techniques can offer statistically optimal solutions at the group level. However, just as group-level relationships cannot fully replace individual relationships [159], solutions appropriate for groups cannot be entirely applied to individuals. Behavioral sciences use longitudinal data collection to better separate within-person and between-person effects [160]. Recent advancements in LLMs have enabled personalization by recording detailed interaction histories and preferences, achieving alignment with users [161]. In human-computer interaction, personalized affective computing models have also emerged, leveraging techniques like fine-tuning and multi-task learning [162]. Therefore, modeling individual affective concerns using these technologies represents the next critical goal for affective agents.

## 4. TELEOLOGY-DRIVEN AFFECTIVE COMPUTING

In Section 3, we reviewed the major research directions and significant achievements in the field of affective computing, along with a detailed analysis of the theoretical foundations behind these advancements. We identified that these studies lack explicit standards and principles for affective interaction at the theoretical level. Hence, in this section, we first propose design principles for interactive affective agents based on teleological perspectives — aligning with personal/individual and even group/collective well-being over extended time scales. Subsequently, we discuss the database and algorithmic models required to align with these principles. The integration of these database and models can help capture the complexity of affective dynamics, revealing the multidimensional characteristics of emotions and their causal relationships. Ultimately, this framework aims to construct a comprehensive teleological affective computing paradigm, transitioning from theoretical principles to corresponding data and algorithms (see **Figure 1**).

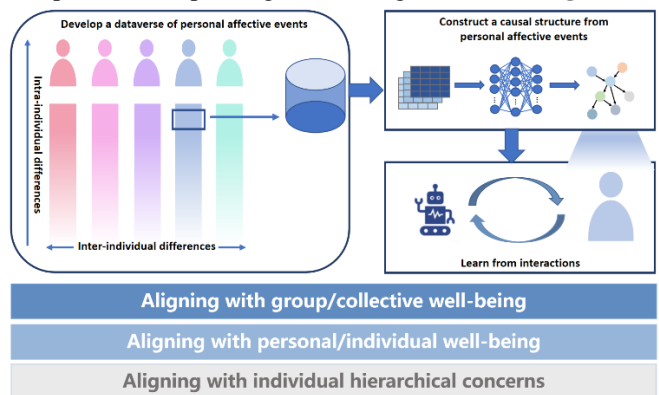

**Figure 1. Overview of the teleological affective computing framework.** This research framework comprises response principles, dataset construction, and algorithm models. Based on two core response principles (highlighted in blue) — aligning with personal/individual well-being and group/collective well-being — the process begins by collecting individual-level affective event data to construct a personal affective event "dataverse." On this basis, causal inference is employed to establish generative models of personal affective events. These models are then used to train agents in virtual environments for effective actions. Ultimately, through repeated meta-reinforcement learning simulations of human-agent interactions in virtual settings, agents dynamically adapt and optimize their alignment with both individual and group/collective well-being.

*4.1 Principles of Affective Interaction Guided by Teleology*

As highlighted in Section 2, the predominant emotion theories converge in advocating teleological principles from different perspectives. Schiller et al. [24] describe this principle as a comprehensive synthesis of existing emotional theories. They characterize emotions as phenomena where actions are performed based on an individual's comfort zone (addressing affective concerns) while monitoring the adaptation process (displaying affective features). For participants in affective interactions, understanding each other's affective concerns is even more critical than identifying immediate affective states. Affective concerns represent the causes, explanations, and relationships between individuals and physical or psychological objects in their environment. Interaction participants can only influence or modulate the other's emotional experience effectively by understanding the objects that underpin their affective concerns. These concerns are structured hierarchically, starting from an individual as the reference point, organized from proximal and specific to distal and abstract levels [24]. However, the hierarchical nature of affective concerns still fails to explain trade-offs among different concerns over extended time scales, raising questions about how individuals define their overall optimal state.

To address this, Schiller et al. [24] propose a global concern algorithm to summarize hierarchically structured concerns at broader temporal scales. Global concerns transcend specific objects, encompassing all object-specific concerns — frequently referred to in affect research as personal well-being [52], [56], [110], [129], [163], [164]. This construct represents the synthesis and prioritization of hierarchically structured concerns, guiding individuals in choosing among multiple competing concerns.

Notably, fulfilling hierarchical concerns does not always contribute to global concerns. For instance, gambling can provide excitement (a heightened physiological pleasure) for some players, sustaining long-term participation. However, as with all addictive behaviors, such prolonged engagement sacrifices other hierarchical concerns. Gamblers might spend excessive money, neglect relationships, or abandon academic and career goals. Even if gambling yields temporary positive experiences, this state is unlikely to be termed "well-being." Becker and Bernecker [165] explored the relationship between immediate pleasures and long-term goals in self-control research, noting that pursuing hedonistic activities often incurs opportunity costs, limiting enjoyment.

The conflict between global concerns and hierarchical concerns embodies tensions between the whole and its parts, long-term versus short-term, and differing levels of positivity. S. Yu [166] suggested that higher-level well-being is not merely the sum of positive states. Temporary negative emotions can create conditions for overall positivity. Particularly when pursuing higher-level affective concerns, individuals often endure complex actions and environmental interactions. During these processes, setbacks might evoke short-term negative emotions that lay the groundwork for subsequent positive feedback. For example, researchers experiencing academic setbacks may feel frustration and disappointment, yet these emotions can prompt critical reflection, improving strategies and leading to valuable insights. Negative emotions, under specific circumstances, do not impede long-term development but foster higher-level efforts and overall positive transformations [166]. Our ultimate pursuit is not moment-to-moment positive experiences but optimal affective states across extended time scales. Trajectories aligned with global concerns lead to positive mood, while deviations foster negativity. When all energy aligns efficiently with global concerns, the organism achieves optimal adaptation to its environment, experiencing a state of effortless control and order, referred to as "flow" [167], [168].

The complex dynamics of positive and negative transitions in affective phenomena motivate a dynamic perspective defined by the multilayered interconnectedness of human experiences [166], [169]. Active inference provides a valuable framework for interpreting these dynamic processes. Its foundational principle involves comparing the brain's internal models with the external environment to generate predictions, which are then iteratively adjusted based on actual feedback. This dynamic feedback regulation enables individuals to minimize cognitive errors and optimize responses to environmental stimuli by updating beliefs and behaviors [170], [171]. Under the active inference perspective, the valence component of emotion is modeled as the rate of error dynamics. Slower-than-expected error reduction signifies negative valence, while faster-than-expected error reduction indicates positive valence [172], [173], [174], [175]. Expanding this concept, Miller et al., [175] distinguished between local and global error dynamics to explain the difference between transient and enduring happiness. Local dynamics involve adjustments to specific action strategies, reflecting task-specific performance in reducing predictive errors. Global dynamics concern an individual's overall predictive performance across multiple life domains, reflecting how uncertainties are managed over a lifetime. To maintain mental well-being, individuals must balance local and global error dynamics, optimizing resource allocation between maintaining current strategies and seeking new ones. While reducing predictive errors within specific domains is crucial, individuals must also address multiple affective concerns across their lives, reallocating resources to sustain growth and adaptability. Through this balancing act, individuals adapt to environmental challenges while introducing temporary uncertainties to foster growth and skill acquisition.

Based on the multilayered analysis of affective phenomena, we can define three hierarchical principles for interactive responses between affective agents and humans (see **Figure 2**): 1) **Aligning with individual hierarchical concerns.** The affective agent can calculate the relevance of a particular stimuli to the individual and help the individual reduce prediction errors in specific tasks; 2) **Aligning with personal/individual well-being.** The affective agent is capable of calculating the relevance of multiple objects to the individual, and also flexibly modeling the process by which the individual updates the weights of different hierarchical concerns over a longer time scale, based on contextual changes; 3) **Aligning with group/collective well-being.** The affective agent can simultaneously consider the well-being of all directly or indirectly interacting individuals and seek ways to maximize group/collective well-being. Ideally, the affective agent should be human-centered, providing highly personalized responses. This not only requires the agent to understand the hierarchical concerns of the interacting individuals, but also to guide the individual toward broader global concerns through responses, interactions, and engagement. It can not only recognize an individual's current needs and emotions but also promote the enhancement of both individual and even group/collective well-being over a longer time scale.

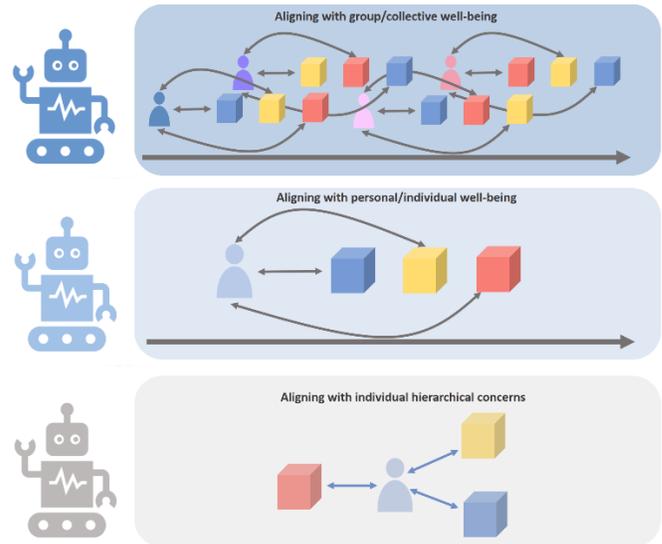

**Figure 2. Three-layered interaction principles under teleological guidance: aligning with individual hierarchical concerns, aligning with personal/individual well-being, and aligning with group/collective well-being.**

The different-colored cubes represent various objects in the environment, while bidirectional arrows indicate relationships between individuals and objects. Aligning with individual hierarchical concerns requires the agent to calculate specific associations between the individual and each object, helping to reduce prediction errors for specific tasks but failing to resolve conflicts between tasks. Aligning with personal/individual well-being involves prioritizing hierarchical concerns in a context-sensitive manner across the time scale represented by the arrows. Aligning with group/collective well-being goes further by simultaneously considering the well-being of multiple individuals, making trade-offs among different interactive objects to maximize group benefits.

*4.2 Develop a "Dataverse" of Personal Affective Events*

Breakthroughs in deep learning owe much to the availability of large-scale datasets [176]. Similarly, constructing computational models that align with well-being requires first collecting data that reflect human growth and well-being. As discussed earlier, teleological perspectives on affects consider affective phenomena as arising from adaptive interactions between agents and their environments. This concept resonates with ecological psychology, which examines how individuals' psychology and behavior are shaped by natural and social environments over extended time scales. The roots of ecological psychology can be traced to James Watson's behaviorism, which emphasized the connection between micro-environments and behavioral responses [177]. In developmental psychology, Bronfenbrenner, influenced by Lewi's field theory and Vygotsky's sociohistorical perspective, proposed an ecological systems theory that systematically analyzed the nested structures of environments individuals inhabit [178], [179], [180]. Building on these foundations, psychologists from diverse domains — cognition, development, motivation, and personality — introduced finer-grained concepts and theories to describe human-environment interactions. Schiller et al. [24] argue that agents in interactions actively orient toward their concerns, which define the content of affective experiences. Dweck [181] posited those three fundamental psychological needs — acceptance, predictability, and competence — underpin motivation and personality, significantly influencing personal well-being and growth, particularly in early life. Extending this, Dweck identified four needs — trust, control, self-esteem/status, and self-coherence — arising from combinations of basic psychological needs. These often demand complex schemas and metacognitive skills. Comparing Schiller et al. [24] and Dweck [181], we find that Schiller's operational concerns align with Dweck's needs for trust, control, and self-esteem/status. Self-coherence, a synthesis of these needs, forms the core of all concerns. By monitoring self-coherence, individuals can closely track their well-being, consistent with Schiller's concept of global concerns (see **Figure 3**). Thus, adaptation to the environment can be seen as agents balancing different concerns or needs based on environmental affordances.

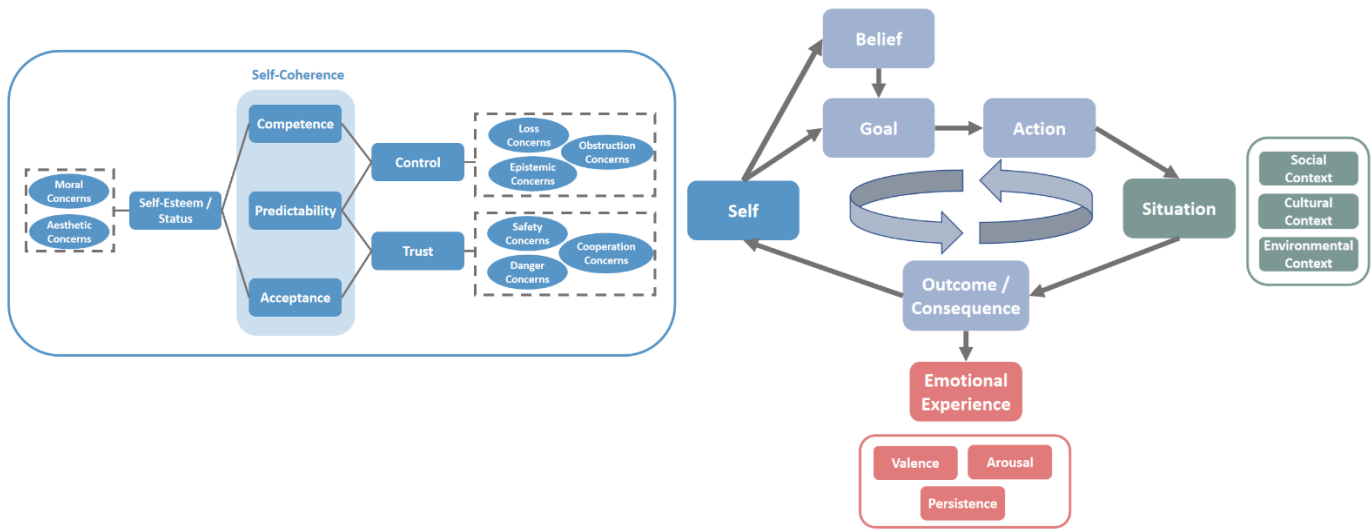

**Figure 3. Structural model of affective events.**
The generation of affective events can be described using the motivational framework proposed by Urhahne and Wijnia [184]. The central circular arrows represent the dynamic interaction between the self and the situation. The model starts with the "Self," which interacts and adapts to the situation through beliefs, goals, actions, and outcomes/consequences to generate specific affective experiences. The blue box on the far left contains components defining an individual's unique basic needs or concerns. According to Dweck's [181] basic need theory, three core needs — competence, predictability, and acceptance — combine to form self-coherence. Their pairwise combinations further represent more complex psychological needs: self-esteem/status, control, and trust. The concerns outlined in Schiller et al. [24] align well with these psychological needs, as the dashed box indicates. Inspired by Bronfenbrenner's [178] ecological systems model, the green box on the far right describes interactive situations more precisely through social, cultural, and environmental contexts. Finally, the red box at the bottom represents the features of affective experiences, including valence, arousal, and persistence.

Although humans share universal needs, individuals differ in the emphasis placed on these needs. For instance, some prioritize self-esteem and social status, seeking recognition and achievement, while others value trust and security, favoring stable relationships and environments. These variations also extend to how individuals fulfill these needs. To achieve need-related goals, individuals must hold beliefs supporting successful goal attainment, shaped by past interactions with their environments [181]. As individuals develop, their needs may evolve throughout life, further amplifying individual differences (see **Figure 4**). Considering these individual differences, it becomes crucial to build personalized affective models. This process requires moving beyond aggregated group data to individual-level data collection and modeling. Abstract statistical summaries (e.g., means or prototypes) often fail to capture individual uniqueness adequately [182]. Research in education by Saqr et al., [183] highlights the limitations of models trained on aggregated data for individual-level predictions, emphasizing the importance of collecting and analyzing individual data.

Differences in individuals' dependency on various needs define each unique self, serving as the motivational starting point for behavior [184]. Needs and concerns shape key domains for personal/individual well-being and optimal development — the "comfort zone" — while also providing the motivational drive to fulfill these needs [181], [185]. The integrated framework proposed by Urhahne and Wijnia [184], which differentiates dimensions of motivation such as self, goals, actions, outcomes, and consequences, illuminates how individuals interact with their environment based on needs, generating emotional experiences.

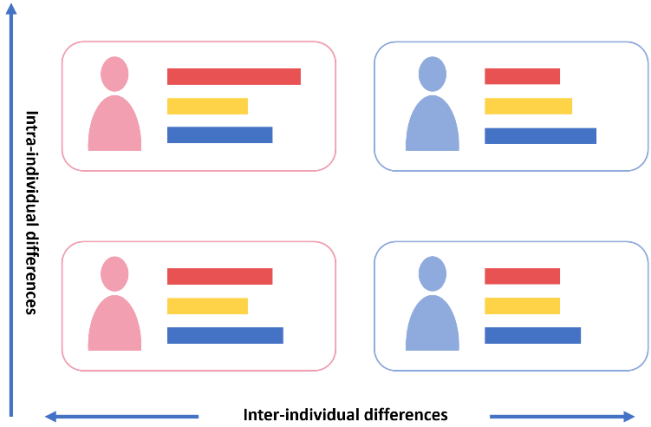

**Figure 4. Inter-individual and intra-individual differences in hierarchical affective concerns (expectations for psychological need fulfillment).** Horizontal comparisons illustrate variations in the composition of psychological needs across individuals, representing inter-individual differences. Vertical comparisons show how an individual's need composition changes across different stages of growth and development, representing intra-individual differences.

Synthesizing Schiller et al.'s explanation of affective phenomena [24], Dweck's basic need theory [181], Bronfenbrenner's ecological systems theory [178], and Urhahne and Wijnia's motivational framework [184] allows for a finer-grained analysis of affective generation processes (see **Figure 3**). When analyzing individual affective experiences, beyond considering valence and arousal, we introduce the dimension of persistence. While most emotion studies focus on momentary intensity, variations in affective persistence offer critical insights into long-term affective impacts [186]. Neuroscientific studies suggest that affective experiences can "spill over" to other stimuli due to sustained amygdala activity following affective arousal [187], [188]. Affective persistence is closely linked to individual health and well-being [189]. Thus, incorporating persistence into affective experience analysis helps identify emotionally impactful events in daily life.

By integrating these diverse theoretical perspectives, we can better describe emotionally significant events for individuals and design corresponding data collection methodologies. Each data entry should provide a comprehensive description of the affective event: the context in which the individual acted, the beliefs guiding their choices, the goals they pursued, the actions they undertook, the outcomes they achieved, and the extent to which these outcomes satisfied or failed to satisfy their internal concerns or needs, ultimately eliciting specific affective experiences. Within this theoretical framework, each variable can exhibit multiple dimensions rather than being limited to scalar values with two extreme endpoints.

These affective event datasets, collected on an individual basis, essentially sample the life histories of individuals. Each small data subset represents an individual's adaptation process to the environment over a specific period. These fragments form a large-scale database of affective events that captures trajectories of affective phenomena across countless human individuals and other affective agents in not only various contexts but also documents affective dynamics at the group level (see **Figure 5**). This group-level entity is not a simple aggregation of individuals but a unique "whole" with distinctive psychological characteristics and affective concerns [190]. In fact, the data used to train models is far more critical than often realized. Muttenthaler et al. [191] demonstrated that the scale and architecture of neural networks have a smaller impact on the consistency between model representations and human behavior than the training dataset and objective functions. Similar to how ImageNet [192] advanced breakthroughs in visual recognition tasks through large-scale, high-quality datasets in computer vision, this affective event "dataverse" serves as the foundation for affective computing, offering immense potential. It can directly address the question central to AI: "What should we learn?" Unlike static datasets constructed in laboratory settings, these affective event datasets must be captured dynamically in real-life situations to authentically reflect the affective adaptation processes of individuals and groups.

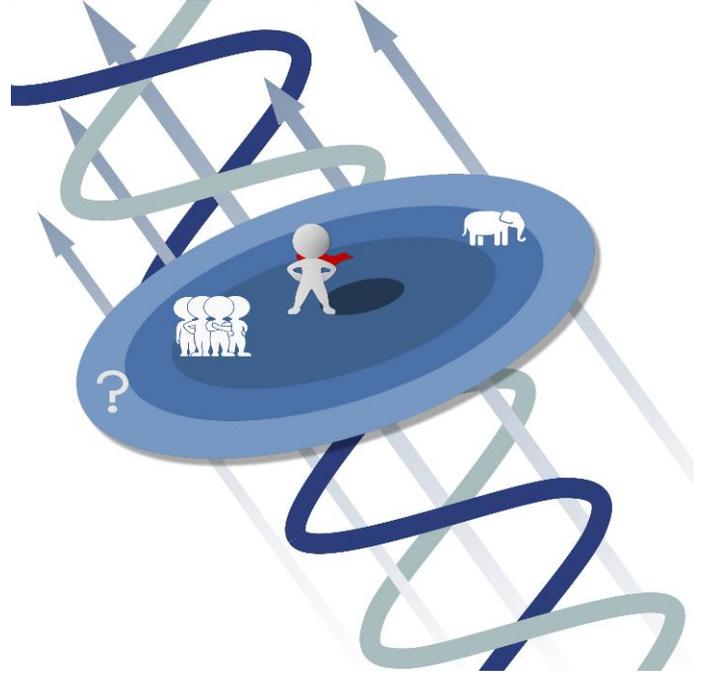

**Figure 5. Schematic representation of the affective event "dataverse."** The blue concentric circles represent the horizontal scope of the affective event "dataverse," encompassing the affective trajectories of individuals, groups, and cross-species affective agents. The spiral arrows symbolize the temporal evolution of affective events, illustrating the longitudinal dynamics of affect. These fragments of affective events converge into a large-scale affective event "dataverse," capturing the affective trajectories and dynamic adaptations of diverse entities in multifaceted contexts.

In this context, dynamic assessment methods provide effective tools for capturing people's momentary states in everyday life [193], [194], [195]. Representative approaches include Ecological Momentary Assessment (EMA) [194] and Day Reconstruction Method (DRM) [196]. EMA uses randomized sampling multiple times a day to capture the dynamic changes in affective events at high frequency, including contextual features, belief states, and immediate reactions, while minimizing recall bias. This real-time collection method not only records the interaction process between individual goals and contexts but also reveals patterns in how affective experiences dynamically evolve with contextual changes. In contrast, DRM provides a more macroscopic perspective, requiring participants to reflect on and document major affective events over the course of a day. This method is suitable for capturing the trajectories of significant affective events and helping individuals summarize their satisfaction of personal needs over longer time scales. Through DRM, researchers can analyze how multidimensional features of affective events intertwine over time and track the realization of affective concerns. Combining these two approaches enables a multi-level data collection strategy that captures both micro-level details of momentary states and macro-level trends of affective events. The complementary nature of real-time and retrospective methods allows for a more precise mapping of the complex structure of affective events.

Beyond sampling and observing individuals' everyday states, Immersive Virtual Reality (IVR) technology offers new possibilities for collecting intervention data on affective events. Currently, IVR technology is widely applied in education [197], [198], [199] and memory research [200]. By constructing highly immersive virtual environments, IVR enables researchers to precisely control situational variables and observe individuals' affective responses under different conditions. This is particularly useful for studying affective events in real life that are difficult to capture or cannot be repeated [197], [201], [202] — such as dangerous situations, rare affective experiences, or high-pressure decision-making processes. In virtual environments, researchers can create scenarios such as public speaking, social conflicts, or emergency rescues while manipulating environmental factors to explore how these elements influence affective events' occurrence and development. Additionally, IVR can elicit affective responses closely resembling real-world reactions, enhancing the ecological validity of affective data. For instance, simulating failure or success in a virtual setting allows researchers to directly observe how changes in goal achievement or need satisfaction affect individuals' affective states.

However, whether using dynamic assessment or IVR technologies, obtaining user data requires addressing privacy and security concerns. Unlike traditional machine learning methods that rely on centralized databases, federated learning allows multiple users (clients) to collaborate in training a shared global model without sharing local device data, thereby protecting user privacy while enabling cross-device collaboration [203]. Additionally, anonymization remains a primary method for reducing privacy risks in data sharing, ensuring that anonymized information cannot be traced back to individuals, thus posing no threat to user privacy [204].

Apart from privacy issues, data bias is another critical challenge. Data collected by researchers for various purposes inevitably carries inherent biases. Increasingly, philosophers of science acknowledge that the ideal of purely value-neutral science is unattainable, as all scientific practices inevitably involve value judgments [205]. Although bias is not a new issue, data-driven AI models may amplify existing biases or even create new ones [206]. Affective event data collected via dynamic assessment and IVR technologies may similarly be prone to biases. In dynamic assessment, biases often arise from participants' memory distortions, which are influenced by selective recall and temporal factors. In IVR, biases may result from discrepancies between virtual scenarios designed by researchers and real-world contexts, potentially impacting participants' affective responses and experiences. To mitigate these issues, researchers can increase the diversity and authenticity of scenarios and use repeated sampling and recall to reduce memory distortions in data. Moreover, enhancing the realism of virtual scenarios and participants' immersion ensures that scenarios are more closely aligned with real-life experiences, reducing biases introduced by virtual settings.

Finally, to implement this data collection approach across broader populations and contexts, we recommend small-scale pilot studies in volunteers' everyday environments. These pilots can address key issues in data collection, such as the effectiveness of scenario design, bias control, and privacy protection measures. After resolving these challenges, the approach can be scaled to more diverse populations and contexts, enabling large-scale, generalized affective event data collection. This iterative process allows for continuous optimization of data collection methods, enhancing data quality and providing richer, more accurate training data for subsequent model development.

*4.3 Constructing Affective Agents Aligned with Personal/Individual and Group/Collective Well-Being*

In addition to high-quality datasets, designing algorithmic architectures for training affective agents aligned with personal/individual and group/collective well-being is equally critical. We define "alignment with well-being" as an agent's capacity to modify the external world through actions to better reflect personal/individual or group/collective preferences. In cognitive psychology terms, this goal requires the agent to possess a theory of mind, enabling it to infer the mental states of others. This overarching objective can be decomposed into two key sub-tasks: (1) **Inferring the causal structure underlying individual affective event data to accurately simulate real individuals with unique comfort zones**; and (2) **utilizing this simulator to interact with affective agents, allowing them to learn, through environmental feedback, how to align personal/individual or group/collective well-being over the long term**. In this section, we explore how to construct affective agents capable of aligning personal/individual and group/collective well-being by addressing these two tasks and their interrelations. Notably, the goal is to clarify challenges and directions rather than provide a fully developed implementation plan.

*4.3.1 Generative Models for Real Individuals: Causal Modeling of Personal Affective Events*

Humans typically understand a specific person's psychological state and predict their responses in future situations by observing and interacting minimally. This is done in an unsupervised manner, independent of any specific task, especially as relevant affective event experiences accumulate over time. For example, when reading Hamlet, although the plot does not explicitly state it, readers can infer from Hamlet's actions that he feels anger and sadness due to the death of his father and the remarriage of his mother. Additionally, we can make counterfactual inferences: if Hamlet's father had not died or if his mother had not married his uncle, Hamlet's affective state might have been entirely different. Through this, we can quickly grasp the hidden causal relationships behind affective events of characters, even with limited information, and predict possible affective changes. As affective experiences accumulate, we are able to effectively infer how an individual will respond emotionally in specific situations and determine whether these affective experiences will persist. However, this task is challenging for machine learning models based on statistical approaches. Most of the success of machine learning is attributed to large-scale pattern recognition on independently and identically distributed (i.i.d.) data, which establishes dependencies between variables [207]. In the field of affective computing, it can model the correlation between externally observable information and an individual's internal affective experience. For example, how likely is a specific facial expression to indicate happiness? Given the observed weather conditions and activities, a person is engaged in, what is the probability that they are experiencing a particular affective state? While these association-based predictions can achieve a reasonable level of accuracy, they may still fall short of truly guiding affective interactions. In the aforementioned example, a facial expression recognition model can determine that a person's expression shows anger, but if we ask, "If this person were smiling instead of frowning, would they still be angry?" the model would typically be unable to answer. Even if it can tell us the correlation between "anger" and "frowning," it cannot predict how the affective state would change if the facial expression changes. This is because statistical models are accurate only under certain assumptions and within the same experimental range, and applying interventions changes the data distribution, leading to inaccurate predictions [207], [208], [209], [210].

Simply learning statistical correlations between variables does not generalize well to non-i.i.d. data; it requires the learning of underlying causal models [208]. Unlike correlations between data, causal relationships focus on structural knowledge of the data-generating process, which allows for interventions and changes, enabling agents to act in the imaginative space [207]. Recent research by Richens and Everitt [211] indicates that learning causal models of the data-generating process is essential for agents to cope with large-scale distributional changes. Only by recognizing causal relationships between features and labels in training data can agents adapt to changes in covariates and labels. In fact, a key characteristic of human cognitive ability is inferring underlying patterns and structures from sparse data [212], and causal structure learning is considered a pathway to achieving AI similar to human cognition [213]. In the field of affective computing, this means we need to build models about individuals from affective event data to faithfully simulate real individual responses in various changing contexts.

For a long time, psychologists have attempted to identify a few key variables that lead to specific affective experiences and explore the possible causal relationships between these variables through experiments. In this process, other variables that may have weaker correlations are often treated as environmental noise. When affective instances emerge from a complex causal network full of relevant meaning, this approach can lead to an explosion of potential variable combinations [182]. Let's imagine the specific process of affective generation. When hiking in the forest, a sudden sound from the grass triggers your fear response. This emotion is not directly caused by the sound itself but could involve multiple factors. Your prior experience, such as having been attacked by wild animals before, links this sound to danger. The current environmental uncertainty, such as the darkness and silence of the forest, amplifies your unease. Your physiological state, such as tension or fatigue, makes your body more sensitive to the sound, with physiological reactions like rapid heartbeat and sweating further intensifying your fear. Cultural interpretations of forest sounds might also lead you to perceive the sound as an ominous omen. These factors do not simply add up but interact non-linearly, forming a complex causal network, meaning the generation of fear is a dynamic, individualized process rather than a direct result of a single factor [182].

One of the core assumptions of interactionist approaches is that there is a complex relationship between behavior and the context in which an individual finds themselves [214]. Contexts include not only the physical world at specific times and places but also the socio-cultural background. Affective experiences arise from the dynamic interaction between an individual's internal environment and the external context. An individual's internal environment is defined by a set of variables, including needs, goals, and beliefs, all of which combine to form the current state of the individual. Affective event data from an individual's experience is longitudinal research conducted over time in different environments. These observational data violate the assumption of independent and identical distribution, preserving only structural independence, thus effectively supporting the identification of causal structures [207], [208], [215]. However, these direct observations of the real world often do not directly include the causal variables we are interested in; instead, we need to learn the representations of these variables from the raw data. Causal representation learning focuses on learning these variables from raw high-dimensional data [207], The raw data $D$ is transformed into a joint distribution sample X of n causal variables, represented as:

$$X = \varphi(D) \qquad (1)$$

where $\varphi$ is a neural network serving as a nonlinear function, mapping causal variables to their joint distribution samples $X \in R^{n \times d}$, and each causal variable is a $d$-dimensional vector. Due to the diversity of data sources, emotional event data may involve multi-modal information such as text, speech, and video, so a unified representation method must be found for these variables. Recent studies have addressed this issue by embedding inputs from different modalities into a common space where semantic similarity is related to distance [216]. Moreover, some joint representation learning methods allow direct comparison of different multi-modal information [217], [218], [219].

The relationships between variables can be represented as a Directed Acyclic Graph (DAG) based on an adjacency matrix $A$, where the elements of $A$ describe the dependencies and directions between the variables. Causal structure learning aims to infer a weighted adjacency matrix that describes these relationships from $X$. A common approach to achieve this is through Structural Equation Modeling (SEM). SEM provides a framework for establishing causal path models from variables to outcomes by assuming linear or nonlinear relationships between them. Specifically, SEM views each variable as a linear combination of other variables plus a noise term, and the weights and directions of the adjacency matrix are estimated through optimization methods [220]. The mathematical relationship between the causal graph structure and the samples can be described by a linear generative model:

$$X = A^T X + Z = \left(I - A^T\right)^{-1} Z \quad (2)$$

where the noise variable $z$ is used to model exogenous variables that are not explicitly included in the causal structure. Equation (2) indicates that each observed variable $X_i$ can be regarded as a linear weighted combination of its direct causal parent variables, plus an independent noise term $Z_i$. This expression can be further extended to non-linear transformations, as demonstrated by Y. Yu et al. [220], who used parameterized neural networks to perform non-linear transformations on $Z$ and $X$, providing a more general version of SEM.

In the Bayesian inference framework, the learning of generative models aims to infer the parameters of the model through observed data, thereby characterizing the causal structure and generation mechanism between variables. Given the observed samples $X^k$ and the distribution of latent variables $p(z)$, the generative model is learned by maximizing the log marginal likelihood:

$$\frac{1}{n}\sum_{k=1}^{n} \log p\left(X^k\right) = \frac{1}{n}\sum_{k=1}^{n} \log \int p\left(X^k, z\right) dz \quad (3)$$

However, the marginalization integrals involved in this equation are often difficult to compute analytically, so variational inference methods are commonly employed in practice. Variational inference approximates the true posterior distribution $p(z|X^k)$ by introducing a parameterized variational posterior distribution $q(z|X^k)$ and optimizing the evidence lower bound (ELBO), which serves as a substitute for directly maximizing the log marginal likelihood. For a single sample, the ELBO is expressed as:

$$\log p\left(X^k\right) \geq L_{ELBO}^k$$
$$L_{ELBO}^k = -D_{KL}\left(q\left(z|X^k\right) \| p(z)\right) + E_{q(z|X^k)}\left[\log p\left(X^k|z\right)\right] \quad (4)$$

where the first term is the Kullback-Leibler (KL) divergence ($\geq 0$), which measures the closeness of the variational posterior $q(z|X^k)$ to the prior distribution $p(z)$, and the second term measures the reconstruction capability of the generative model under the variational posterior. In generative models, the variable $z$ is used to reconstruct $X^k$ with a probability density $p(X^k|z)$. The adjacency matrix $A$ is a parameter that needs to be learned. By maximizing the $L_{ELBO}^k$, we can efficiently approximate the true posterior distribution and infer the adjacency matrix $A$ in the generative model. To ensure that the learned $A$ satisfies the acyclic property, the loss function often includes an acyclicity constraint [221]. This process effectively addresses the computational difficulty of direct marginalization integrals and provides a feasible approximate inference method for learning complex generative models.

This generative model is capable of simulating real individuals with unique comfort zones and exhibiting their affective experience characteristics under various hypothetical scenarios. Compared to existing human behavior simulation methods based on LLMs [222], [223], the generative model constructed from affective event data introduces longitudinal behavioral data, uncovering the complex interactions between individuals and their environments — interactions that typically exceed the scope of conventional language training corpora. Moreover, by capturing latent intrinsic causal structures rather than relying solely on correlation-based modeling, this model achieves a higher level of simulation accuracy. By applying hypothetical conditions to key nodes, the model can predict specific individuals' behavioral patterns along with their accompanying affective experiences. Hierarchical concerns can be interpreted as combinations of external environmental node states that correspond to different positive experiences.

*4.3.2 Learning from Simulated Interactions: Training Affective agents through Meta-Reinforcement Learning*

Training an agent capable of taking action and engaging in positive interactions with users cannot typically rely on supervised learning, as it is challenging to obtain a large dataset of optimal action decisions, especially when considering different interaction targets and contexts. Thus, the agent must learn what constitutes good actions through continual trial-and-error interactions with its environment — a process known as Reinforcement Learning (RL) [224]. In the classical RL paradigm, an agent interacts extensively with its environment, observes the current state $s$, and generates an action $a$ through a policy $\pi$. This action leads to a change in the environment's state and a feedback reward $r$ from the environment. The goal of the agent's actions is to maximize cumulative reward $G_t$.

Although RL has enabled agents to perform well in narrowly defined tasks, these agents often fail to generalize their capabilities to new tasks [225]. Evidently, we aspire to train a general-purpose affective agent rather than needing to collect vast amounts of data and retrain it from scratch for every new interaction target or context. To effectively handle diverse real-world contexts and interaction targets, an ideal affective agent must be capable of quickly adapting to changes in its external environment after only a few interactions while continuously improving its behavior as more data accumulates. This ability, referred to as continual learning, is fundamental to constructing adaptive AI systems [226].

Building such an adaptive system is critical for training general-purpose affective agents. The complex causal network underlying individual affective event data is not static but evolves dynamically over time as individuals develop. Piaget's genetic epistemology provides strong theoretical support for this viewpoint. Genetic epistemology emphasizes that individuals' cognitive structures (schemas) continuously develop through interactions with their environment. This nonlinear process, characterized by constant assimilation and accommodation, enables individuals to progress from one state of equilibrium to another [227]. Consequently, an individual's affective responses and processing mechanisms dynamically adjust with changes in their cognitive structure, reflecting the ongoing evolution of their cognitive and affective framework. This continual developmental process can also be observed at the group level. Comparing human culture, animal culture, epigenetics, and parental effects, Morgan & Feldman [228] concluded that the uniqueness of human culture lies in its ability to continuously accumulate and never cease developing. This inherent potential for infinite development implies that the generative model for affective event data must account for not only inter-individual differences but also intra-individual variability. The data observed over a given period represents only a fragment of an individual's life trajectory. In addressing the dynamic changes within interaction models and external environments, fostering agents' adaptability may be a more effective strategy than constructing precise global dynamics models (see **Figure 6**).

In typical continual learning setups, the learner must sequentially acquire knowledge across various content, including new tasks, new examples of old tasks, and different contexts, to adapt incrementally to dynamically distributed data [229], [230]. The primary challenge lies in achieving efficient adaptation to new distributions while maintaining the ability to capture prior distributions — a balance between learning plasticity and memory stability [226], [231]. In this context, Khetarpal et al. [232] highlighted the natural alignment between continual learning and RL. The interactive paradigm of RL provides a robust framework for studying the trade-off between stability and plasticity. This interaction mechanism allows dynamic temporal settings to be naturally modeled as continuous environment problems in RL, enabling solutions to continual learning challenges while leveraging RL theories and methods.

Meta-Reinforcement Learning (meta-RL), a representative method in continual RL, offers an effective solution. The core idea of meta-learning is to acquire knowledge across tasks to enable rapid adaptation to new tasks, thereby reducing dependence on large training datasets [233], [234], [235], [236]. By learning how to adapt efficiently across multiple tasks, meta-learning equips learners to quickly adjust strategies for new tasks while retaining memory of old ones. This minimizes the risk of catastrophic forgetting and enhances long-term learning capabilities in dynamic environments [237]. One significant advantage of meta-learning is its ability to imbue neural network systems with inductive biases akin to symbolic models [238]. In the language domain, McCoy & Griffiths [239] demonstrated that meta-learning effectively transfers the strong inductive biases of Bayesian models into neural networks. Using a Model-Agnostic Meta-Learning (MAML) algorithm, researchers were able to integrate Bayesian priors (symbolic grammar) into neural networks, endowing them with the inductive biases of symbolic models. This enabled neural networks to efficiently learn language structures even in low-data settings. Recently, the meta-learning framework has proven to be an ideal tool for constructing general-purpose models [233].

To train general-purpose affective agents capable of adapting to dynamic environments through meta-RL, a key challenge is designing and constructing virtual training environments encompassing diverse interaction targets and contexts. Given the high costs and limitations of acquiring real-world data, RL agents typically rely on virtual environments for training [240].To ensure that agents trained in virtual environments are deployable in real-world scenarios, researchers must design environments that closely simulate real-world physical and social conditions. For physical environments, numerous simulators already exist that replicate real-world scenarios and physical laws (e.g., [241], [242], [243]), providing agents with high-fidelity sensory and action settings. For social environments, modeling must further encompass diverse individual behavior patterns. For instance, Qi et al., [244] developed a simulator named CivRealm, inspired by the game Civilization. This simulator offers an open-ended stochastic environment where agents can learn complex interpersonal interaction rules in diplomacy and negotiation. However, physical-rule-based simulators alone are insufficient for affective agents. Affective agents must handle more complex task scenarios, including understanding and responding to interaction targets' affective concerns and balancing multiple goals in dynamic contexts.

To achieve this, the virtual training environment must be capable of simulating not only the real physical world but also incorporating generative models that emulate the affective dynamics of real individuals (refer to Section 4.3.1). The former provides high-fidelity sensory and action settings, while the latter simulates complex affective interactions at the social level. Affective generative models, trained on real individuals' affective event data, can generate context-sensitive virtual inhabitants whose behaviors and responses are influenced by affective concerns, the environment, and the agent's actions. Embedding these generative models into world-model scripts allows the creation of dynamic 3D environments with diverse affective interaction tasks. Examples include agents collaborating with multiple virtual inhabitants to complete resource allocation tasks, finding balanced solutions in emotionally tense conflict scenarios, or providing comfort and support through verbal and behavioral means. Affective agents must process high-dimensional stimuli, understand the affective concerns of interaction targets, and plan response pathways based on environmental affordances. Additionally, as the generative models are derived directly from real individuals' affective data, these virtual inhabitants exhibit highly personalized affective characteristics, making training tasks more realistic. In such multimodal affective interaction environments, affective agents can incrementally develop capabilities for perceiving, adapting to, and planning for complex affective scenarios through meta-RL, laying a solid foundation for real-world deployment. Research from the machine learning community has already demonstrated the feasibility of training agents capable of performing a wide variety of tasks (e.g., [245], [246]).

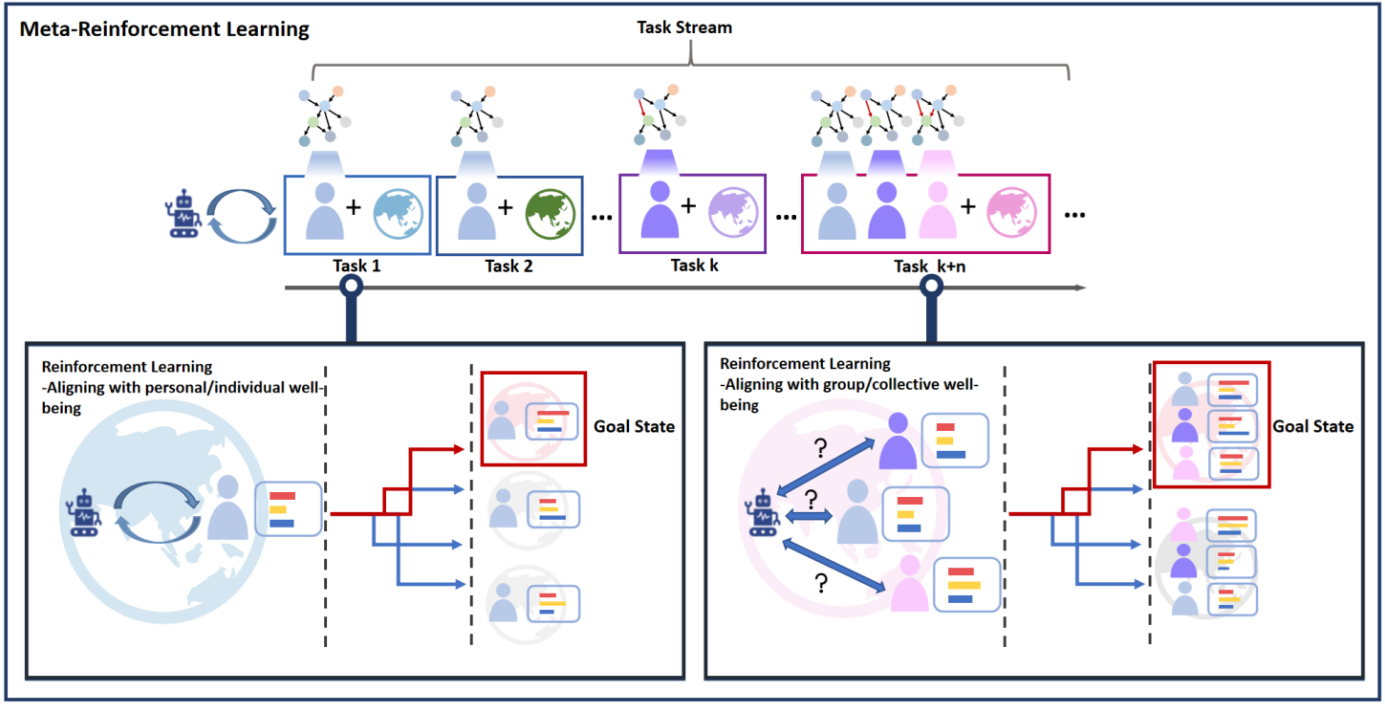

**Figure 6. Training trajectory of affective agents in the Meta-Reinforcement Learning (Meta-RL) framework, progressing from aligning with personal/individual well-being to group/collective well-being.**
The upper section of the figure represents the outer loop of Meta-RL, where the agent gradually adapts across tasks, evolving from individual-level tasks (e.g., interacting with a single user) to more complex group-level tasks (e.g., coordinating affective states across multiple users). Each task in the task flow consists of two components: a generative model and a world model. The humanoid icon represents the generative model, which is built using real-world affective event data to simulate the affective responses and decision-making processes of virtual inhabitants. The globe icon represents the world model, which provides high-fidelity perception and action scenarios to simulate the external environment required for physical and social interactions within the task. The lower-left and lower-right sections represent the inner loops of the Meta-RL process, corresponding to specific implementations of aligning with personal/individual and group/collective well-being, respectively. In the early stages, the focus is on aligning with personal/individual well-being by exploring potential strategies through reinforcement learning (illustrated by the blue pathway), ultimately identifying the optimal strategy (red pathway) that guides the user's affective state toward the target state. In tasks aimed at aligning group/collective well-being (lower right), the agent must address uncertainties such as conflicts in individual preferences and incomplete signals, integrating the diverse needs of multiple users to optimize affective interaction pathways and achieve a harmonious and optimal group/collective affective state. Through Meta-RL, affective agents generalize knowledge across tasks, enhancing their adaptability in dynamic environments and demonstrating their potential for scaling from individual emotion optimization to social collaboration systems.

In RL, environmental rewards are the core mechanism driving an agent's ability to learn complex skills. As noted by Silver et al. [247], reward mechanisms can enable AI to develop various sophisticated abilities, including language comprehension, learning, social interaction, generalization, and imitation. In affective interaction tasks, the design of an appropriate reward function is crucial for training affective agents. These agents need to dynamically adjust their behavior based on environmental affordances to assist interaction partners in achieving their affective concerns and, in doing so, receive rewards. However, in practical scenarios, fully satisfying all levels of affective concerns is often unachievable, especially considering the inherent conflicts between certain layers of affective needs. Therefore, the reward function should assign differentiated weights to various levels of affective concerns to establish clear priorities. This weighting mechanism emulates how individuals integrate multiple goals and optimize resource allocation in the pursuit of long-term well-being. In this process, the three dimensions of affective experience — valence, arousal, and

persistence — play a critical role in shaping the affective agent's behavioral choices. Valence indicates whether an emotion is positive or negative; the agent should prioritize actions that promote positive experiences while mitigating negative ones. Arousal reflects the intensity of an affective state, directly influencing the agent's response urgency and intensity. High-arousal affective states demand immediate responses, while low-arousal contexts may require more measured and cautious behavioral strategies. Persistence, on the other hand, pertains to the duration of an affective experience, influencing how affective goals should be pursued. The reward mechanism should balance short-term rewards with long-term objectives, preventing immediate gratification from undermining long-term well-being.

Achieving global concerns does not necessitate simultaneously fulfilling all levels of affective concerns; rather, it emphasizes the effective allocation of resources to achieve higher-level holistic coordination. While the specific means of achieving global concerns vary among individuals, positive psychology theories provide universal descriptions of well-being, offering principled guidance for reward function design. These theories present two primary perspectives on well-being: hedonism and eudaimonia. The former defines happiness as the accumulation of positive experiences [248]. Although satisfying affective concerns at different levels can bring positive experiences, the quality and longevity of these experiences vary. Within the hedonistic tradition, goals that yield longer-lasting positive experiences should be prioritized with higher weight. Conversely, the eudaimonic perspective views a fulfilling and deeply satisfying life as the true definition of well-being, emphasizing the realization of personal potential [249]. From this viewpoint, affective agents should not only help achieve immediate needs, such as fulfilling physiological demands, but also actively facilitate goals that are distant in metabolic impact, such as exploring the environment to satisfy curiosity. The latter often contributes more to a fulfilling and meaningful life. In this context, highly aroused states (e.g., anxiety or feelings of urgency) indicate the need for affective agents to respond swiftly and adaptively to prevent extreme affective deprivation.

We propose that these seemingly opposing views on well-being — short-term gratification and long-term fulfillment — may represent different aspects of the same overall well-being state. Achieving short-term, intense positive affect in a rigid, addictive manner is unsustainable, as it hampers an individual's ability to explore environmental affordances, leading to fixation on specific attractors [166], [175], [250]. In contrast, goals that require complex actions and delayed feedback often lead to richer and more enduring positive experiences, such as shaping the environment. Similarly, pursuing self-actualization does not imply the rejection of all pleasurable activities; for instance, prolonged mental and physical exhaustion is detrimental to creative endeavors.

A feasible approach, integrating the aforementioned perspectives, is to assign higher priority to affective concerns that have a distant metabolic impact yet contribute to long-term positive experiences, while dynamically adjusting weight allocations based on changes in arousal levels. This ensures that pressing affective concerns are not left unmet for extended periods, thereby achieving a dynamic balance between pursuing complex goals and maintaining fundamental stability (see **Figure 7**). This approach is further supported by the Tri-Reference-Point theory proposed by X. T. Wang and Johnson [251]. The tri-reference-point theory posits that decision-makers typically consider three critical points when making choices: the current state, the optimal goal, and the worst-case scenario [251]. In affective computing, the current state ensures that urgent affective concerns are promptly addressed, the optimal goal focuses on achieving long-term affective equilibrium, and the worst-case scenario prevents the system from becoming imbalanced due to prolonged neglect of specific affective concerns. By dynamically adjusting weight allocations, the system can effectively balance short-term demands and long-term objectives. When an affective agent detects that the interaction partner is in a high-arousal affective state, it indicates an urgent and intense affective concern. In such cases, the agent must take immediate action to address the present affective concern, ensuring timely and prioritized responses to high-arousal states. This responsive mechanism enables the affective agent to remain sensitive to the user's current affective state while simultaneously working toward long-term well-being, thereby optimizing the interaction experience dynamically. For instance, when an individual experiences a positive valence and high-arousal emotional state — such as joy, excitement, or a sense of achievement — they exhibit heightened responsiveness to environmental stimuli. In this scenario, the affective agent should adopt an encouraging and supportive approach, reinforcing the user's positive emotions and further facilitating goal attainment. For example, when an individual completes a task and feels excited, the affective agent can enhance this positive emotion by offering praise, rewards, or new exploration opportunities, thereby enriching the experience and motivating the individual to pursue higher achievements. This strategy not only helps maintain positive emotions but also fosters deeper engagement in the interaction process. Conversely, when an individual experiences a negative valence and high-arousal emotional state — such as anxiety, anger, or distress — the affective agent must employ emotional soothing strategies to help the individual regain emotional stability. For instance, if a person feels anxious about a particular issue, the affective agent can guide them through deep breathing exercises, provide emotional support, or offer reassuring language to alleviate distress. Additionally, the agent can assist in identifying the root cause of the negative emotion and suggest coping strategies or psychological support to help the individual manage their emotions effectively. By implementing such strategies, the affective agent not only alleviates negative emotions but also promotes emotional self-regulation, helping individuals regain composure and stability. This, in turn, enhances the overall affective interaction experience, ensuring that both immediate needs and long-term well-being goals are addressed in a balanced and adaptive manner.

This design logic can also be extended to affective interaction scenarios involving multiple interaction targets, simulating how agents optimize group/collective well-being. Within this framework, the definition of global well-being can expand from focusing on a single individual to encompassing the joint well-being of all interaction targets, thereby supporting affective agents in aligning overall interests within multi-agent ecosystems. The range of interaction targets can go beyond human agents to include all living beings in the natural world, aiming for the maximization of ecosystem well-being. However, addressing the prioritization of well-being across different interaction targets poses a complex issue involving ethics and social values, which requires further exploration. In summary, by scientifically designing reward functions, affective interaction principles can be effectively embedded into the action objectives of affective agents, providing both theoretical and practical support for deep collaboration between humans and algorithms.

Beyond creating a large number of complex training tasks, to ensure the learning progress of affective agents, we must also design a training strategy that allows the agent to sample tasks within its "zone of proximal development" [245], thereby improving agent performance and sample efficiency. The core of this strategy lies in controlling the dynamic changes in the environment, which can be reflected in two aspects: first, changes in task competency requirements, and second, changes in the internal structure of interaction targets. Regarding task competency requirements, the difficulty of training tasks should transition smoothly and gradually increase, rather than abruptly jumping to tasks far beyond the agent's current capability, ensuring that the agent can continually enhance its ability to respond within a relatively stable capability range. As for changes in the internal structure of interaction targets, the affective needs or behavioral characteristics of virtual environment interaction targets should not change too drastically but instead should align with the learning stages of the agent, adjusting incrementally. By controlling the gradual changes in task competency requirements and the internal structure of interaction targets, the agent can form a stable learning path between adjacent tasks, thereby gradually mastering more complex coping strategies in affective interactions and simulating the natural process of individual development. This strategy not only accelerates the agent's learning efficiency but also improves its adaptability and generalization capabilities in dealing with various affective concerns and interaction needs in complex and dynamic real-world environments.

Overall, meta-RL is an ideal tool for building general-purpose affective agents. It effectively promotes the rapid adaptation and continual learning of agents in dynamic environments. By constructing structural models based on individual affective event data, we can design diverse and challenging virtual training environments, driving agents to continually optimize their affective response strategies.

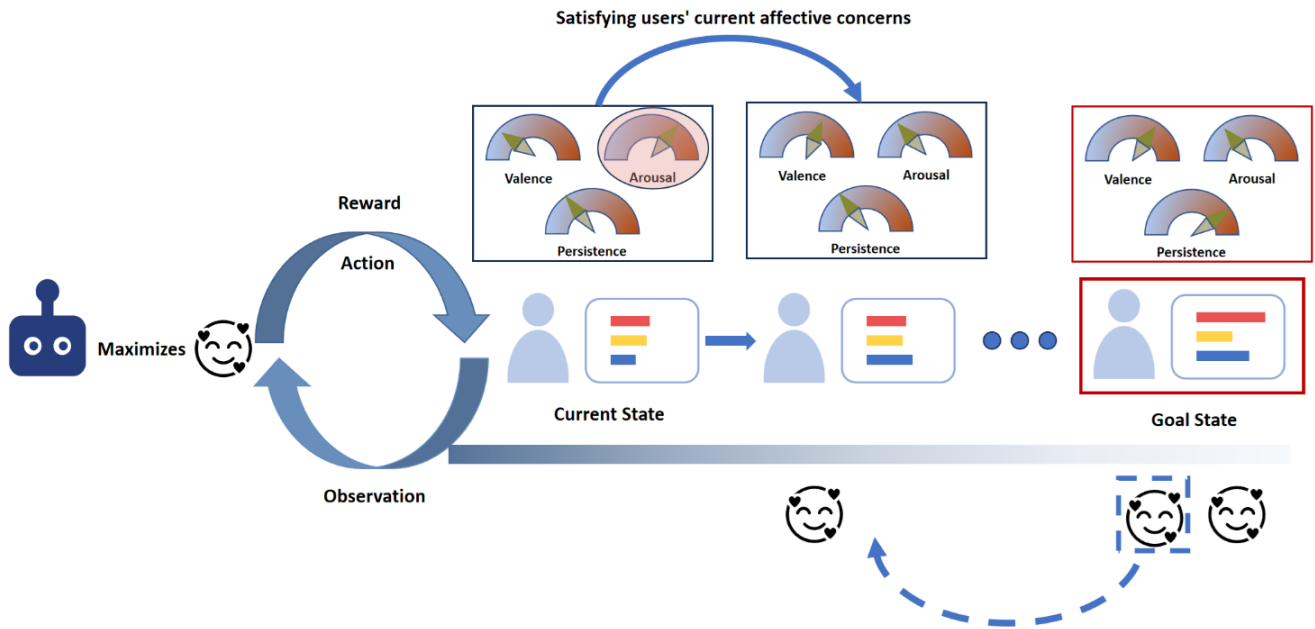

**Figure 7. Reward function design for affective agents to balance long-term goals with immediate affective concerns.**
During interactions with users, affective agents take actions to maximize rewards, represented by smiley icons. The agent aims to fulfill various affective concerns of the user, guiding them toward a target state characterized by positive valence, moderate arousal, and high affective persistence — indicative of long-lasting positive experiences. However, while pursuing long-term goals, the agent must also remain responsive to the user's pressing affective concerns in the present moment. When a user exhibits negative emotions accompanied by high arousal levels, it signals a significant unmet affective need. In such instances, the reward weighting dynamically adjusts to prompt the agent to take immediate actions that address the current affective deficit. This adaptive approach ensures an optimal balance between addressing short-term affective concerns and achieving overarching long-term well-being goals.

## 5. CONCLUSION

In this paper, we present a research framework based on the fundamental principles of teleology aimed at guiding agents' behavior in perception, decision-making, and action. The criterion for aligning with long-term well-being requires agents to more carefully identify subtle differences between individual positive experiences and prioritize helping individuals pursue goals that bring sustained positive experiences rather than short-term, high-arousal intense stimuli, which may have addictive properties. Decoding the internal states of others and understanding what truly benefits long-term well-being are key steps in developing agents' empathy. On this foundation, enabling AI to empathize with the harm or joy of others may require AI to possess various affective concerns that can be fulfilled by the environment [252], as well as instructions to simulate and maintain internal homeostasis [253].

However, although affective agents aligned with long-term well-being provide an ideal path for achieving personal/individual well-being, they still face the risk of bias when considering group well-being alignment. A group consists of a set of interrelated, interacting "elements," and what benefits the group may not necessarily maximize the well-being of each group member [166]. Thus, balancing the well-being of each "element" when considering the "whole" becomes an important and inevitable non-technical issue. In fact, sympathy and care in human society are always biased [254], [255]. Increasing evidence shows that people are more likely to empathize with individuals who resemble themselves [256], [257]. This bias is especially prominent in the alignment of group well-being. How to design systems that can fairly balance the needs of both the group and the individual remains a challenge to be addressed.

In this paper, we did not provide a direct and explicit answer to this issue. Instead, we propose a reliable path for developing scalable capabilities — large-scale simulations based on meta-RL. By incorporating basic social principles into the design of reward functions, agents can plan optimal paths within limited attempts, potentially benefiting human well-being at a level beyond current human wisdom. In this process, we advocate for integrating more psychological and sociological theories, particularly the discussions of well-being in positive psychology. We suggest that principles aimed at helping others pursue more lasting positive experiences should be incorporated into the design of reward functions. AI should prioritize goals that bring sustained positive experiences to individuals, such as building strong interpersonal relationships or achieving self-fulfillment and spiritual satisfaction, rather than pursuing short-term pleasure or avoiding negative emotions. However, whether such long-term, profound affective experiences are more beneficial for long-term well-being or if other answers exist remains an open question that requires further research by psychologists. In addition, directly modeling the developmental process of core needs from infancy to old age [181] may provide a more complete approach than training adaptive models. For example, well-being in infancy primarily relies on basic physiological needs and dependence on caregivers, while, as individuals age, their well-being needs gradually shift toward higher-level self-actualization and social connection. In this process, agents need to dynamically adjust their understanding of well-being and action strategies based on the individual's life stage and psychological development needs to better support long-term well-being. Finally, while we emphasize personalized modeling and response, we do not exclude existing open statistical techniques, which may provide important priors to help agents better understand affective dynamics and individual differences. The semantic space theory proposed by Cowen & Keltner [258] can help capture the systematic changes in emotion-related behaviors, and studies show that emotional patterns have complex, multidimensional characteristics and exhibit some common regularities across different cultures and situations, although their specific manifestations may vary depending on cultural and environmental differences [259], [260]. This line of research can provide strong empirical and methodological support for how we can use group-level statistical data to gain prior information on affective cognition.

In conclusion, in meta-RL-based simulations, agents can adjust according to these psychological and sociological theories, while also conducting large-scale simulations to explore how to balance the complex relationship between individual and group well-being in diverse social environments. This simulation process enables agents to continuously adapt and optimize its behavior strategies in dynamic and changing environments, aiming to maximize individual or group well-being. By integrating theoretical achievements from psychology, sociology, and other fields, this approach not only provides a solid theoretical foundation for the design of affective agents but also offers new ideas and methods for applying these theories to complex social contexts. Ultimately, this interdisciplinary integration aims not to provide a "perfect" answer directly but through multi-level simulations and analyses to help us better discover and solve potential problems, thereby promoting more harmonious development between AI and human society.

### ACKNOWLEDGMENT

This work was funded by the Joint Fund to Promote Cross-Straits Scientific and Technological Cooperation (Project No. U1805263) and the Humanities and Social Science Fund of the Ministry of Education of China (Project No. 23YJAZH183). B. Yin is also partially supported by the Research Start-Up Fund for "Overseas Talent—Young Talent" from Fujian Normal University (Project No. Z0210509). The funders had no role in the preparation of this work.

### AUTHOR CONTRIBUTIONS

B. Yin initiated and supervised the work presented in this article and played a leading role in developing the research framework. C.-Y. Liu conducted critical research in relevant areas. Together, B. Yin and C.-Y. Liu created the initial structure of the paper and led discussions on its iterative improvements. Both authors wrote the majority of the sections of the paper. L. Fu and J. Zhang initiated discussions on several

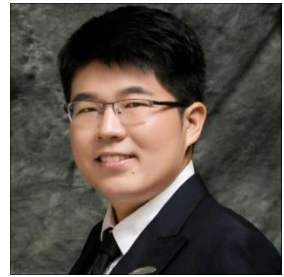

**Dr. Bin Yin** is an associate professor at Fujian Normal University in Fuzhou, China. He was born in China and earned dual Bachelor's degrees in biological sciences and psychology from Peking University in 2008. He later earned his Ph.D. in psychology and neuroscience from Duke University in 2016, where his research focused on animal models and computational methods for understanding reward learning, spatiotemporal cognition, and emotional processes.

Dr. Yin's academic career has centered on affective and behavioral neuroscience and computational psychology, where he integrates experimental approaches with computational modeling to address both fundamental and applied questions. He currently leads research investigating the relationship between preclinical studies and their application to mental health and artificial intelligence. He has published widely in areas such as emotional processing and responses, affective and behavioral development, and affective computing. Dr. Yin is also committed to advancing interdisciplinary collaboration in well-being sciences and is dedicated to contributing to global scientific discussions.

Dr. Yin is a member of the Society for Neuroscience, the American Psychological Association, and IEEE. He has been actively involved in promoting the development of the interdisciplinary research community through various leadership roles. He has received several research grants and awards and has contributed to the editorial process for affective and behavioral science-related publications.

Ms. Chong-Yi Liu and Ms. Liya Fu are master's students mentored by Dr. Yin. Dr. Jinkun Zhang is a colleague and collaborator whose main research interests lie in utilizing insights from basic cognitive research to inform multimedia learning by students.